\documentclass[fleqn,usenatbib]{mnras}
\usepackage{newtxtext,newtxmath}
\usepackage[T1]{fontenc}
\usepackage{caption}
\usepackage{subcaption}
\usepackage{graphicx}
\usepackage{multicol}
\DeclareRobustCommand{\VAN}[3]{#2}
\let\VANthebibliography\thebibliography
\def\thebibliography{\DeclareRobustCommand{\VAN}[3]{##3}\VANthebibliography}

\usepackage{graphicx}	% Including figure files
\usepackage{amsmath}	% Advanced maths commands
\usepackage{cleveref}
\newcommand{\msun}{\,\mathrm{M}_\odot}

\newcommand{\kms}{\,\mathrm{km}\,\mathrm{s}^{-1}}

\newcommand{\au}{\,\mathrm{AU}}

\newcommand{\Myr}{\,\mathrm{Myr}}
\newcommand{\kyr}{\,\mathrm{kyr}}
\newcommand{\yr}{\,\mathrm{yr}}

\newcommand{\pc}{\,\mathrm{pc}}

\title[Simulated Analogues I]{Simulated Analogues I: apparent and physical evolution of young binary protostellar systems}

\author[Tuhtan et al.]{
Vito Tuhtan,$^{1}$
Rami Al-Belmpeisi,$^{1}$
Mikkel Bregning Christensen,$^{1}$
Rajika Kuruwita,$^{2,1}$
and Troels Haugb{\o}lle$^{1}$\thanks{E-mail: haugboel@nbi.ku.dk}
\\
$^{1}$Niels Bohr Institute,
University of Copenhagen, {\O}ster Voldgade 5, DK-1350 Copenhagen, Denmark\\
$^{2}$ Heidelberg Institute for Theoretical Studies, Schlo{\ss}-Wolfsbrunnenweg 35, 69118 Heidelberg, Germany\\
}

\date{Accepted XXX. Received YYY; in original form ZZZ}

\pubyear{2023}

\begin{document}
\label{firstpage}
\pagerange{\pageref{firstpage}--\pageref{lastpage}}
\maketitle

% Abstract of the paper
\begin{abstract}
Protostellar binaries harbour complex environment morphologies. Observations represent a snapshot in time, and projection and optical depth effects impair our ability to interpret them. Careful comparison with high-resolution models that include the larger star-forming region can help isolate the driving physical processes and give observations context in the time domain. We carry out zoom-in simulations with AU-scale resolution, and for the first time ever we follow the evolution until a circumbinary disk is formed. We investigate the gas dynamics around the young stars and extract disk sizes. Using radiative transfer, we obtain evolutionary tracers of the binary systems. We find that the centrifugal radius in prestellar cores is a poor estimator of the resulting disk size due to angular momentum transport at all scales. For binaries, the disk sizes are regulated periodically by the binary orbit, having larger radii close to the apastron. The bolometric temperature differs systematically between edge-on and face-on views and shows a high frequency time dependence correlated with the binary orbit and a low frequency time dependence with larger episodic accretion events. These oscillations can bring the system appearance to change rapidly from class 0 to class I and for short time periods even bring it to class II. The highly complex structure in early stages, as well as the binary orbit itself, affects the classical interpretation of protostellar classes and direct translation to evolutionary stages has to be done with caution and include other evolutionary indicators such as the extent of envelope material.
\end{abstract}


\section{Introduction}

Stars form in the densest parts of molecular clouds \citep{filementary_cores} and evolve through gas accretion from their surrounding environment and dynamical interaction with nearby companions. Due to the high column density, Young stellar objects (YSOs) are predominantly observed at long wavelengths and sorted based on the shape of their Spectral Energy Distribution (SED) into classes \citep{lada1984nature,andre1993submillimeter, 1994ApJ...420..837A,adams1987spectral}. Ranging from class 0 to III, this classification scheme represents the degree of embeddedness and is widely translated into an evolutionary progression for YSOs. First, the collapse of a core embedded in the Giant Molecular Cloud (GMC) produces a protostar. At this stage, the protostar and its disk are embedded in an infalling dusty envelope and depending on the observed dust temperature it is either classified as Class 0 or Class I. When the envelope has been consumed, photons directly emanating from the stellar photosphere can be observed. The newborn star is surrounded by a disk of gas and dust. This corresponds to Class II observations of embedded T-Tauri stars with mid-infrared and near-infrared excesses in the SED.
After the flattening and dispersal of the gas disk, what is left is an isolated pre-main-sequence star with planetary or stellar companions. This corresponds to Class III. \citet{myers1993bolometric} found that the bolometric temperature $T_{bol}$ of an observed system is a good indicator for the shape of the SED and in this paper, we will use their definition of the YSO class in terms of $T_{bol}$
\begin{equation}
\text{YSO Class} = \left\{ \begin{array}{rl}
0 & \text{if }\, T_{bol} < 70 \mathrm{K}, \\
\text{I} & \text{if }\, 70 K \le T_{bol} < 650 \mathrm{K}, \\
\text{II}& \text{if }\, 650 K \le T_{bol} < 2800 \mathrm{K}, \\
\text{III} & \text{if }\, T_{bol} \ge 2800 \mathrm{K}.
\end{array}\right.   
\label{eqn:ysoclass}
\end{equation}

\begin{table*}
\centering
\begin{tabular}{lrrrrrrrrrrrrrr}
\hline
Level           & 7    & 8    & 9     &  10   &  11    &  12  &  13 &  14 & 15 & 16 & 17  & 18 & 19 & 20   \\
\hline
Cell size [AU]  & 6446 & 3223 & 1611  & 806   & 403    & 201  & 101 &  50 & 25 & 13 &  6 & 3.1 & 1.6 & 0.8 \\
$L_{J,min}$& 115  &   58 & 28.8  & 28.8  & 28.8   & 28.8 & 32  &  36 & 36 & 48 & 48 &  64 & 64  & 64  \\
Distance [pc]   &   1  &  0.5 & 0.25  & 0.125 & 0.0625 & 0.04848 & \ldots \\ 
\hline
\end{tabular}
\caption{Criteria for refining a cell at a given level of refinement with respect to the main box-size in the zoom-in runs. The distance of a cell to the primary star is a necessary but not sufficient condition for refinement. A distance of $0.04848\pc$ corresponds to $10\,000\au$. $L_{J,min}$ is the minimum number of cells per Jeans length at the given level of refinement and corresponds to a threshold density for refinement (see Eq.~\ref{eq:LJ}).}
\label{tab:refinement}
\end{table*}

\begin{table}
\centering
\begin{tabular}{lccccr}
\hline
Name & $\Delta x_{min}$ & Final Age & M$_\textrm{prim}$ &  M$_\textrm{sec}$ & Periods \\
     & AU   & kyr      &  $M_\odot$        &  $M_\odot$       &    \\
\hline
\textbf{S1}  & 0.8       & 110          & 0.99 & N/A       & N/A      \\ 
\textbf{B1}  & 3.1       & 148          & 1.07 & 0.34      & 131      \\
\textbf{B1}$^\ast$  & 0.8       & 71          & 0.55 & 0.23     & 27      \\
\textbf{B2}  & 3.1       & 257          & 2.93 & 1.47      & 103      \\
\textbf{B3}  & 3.1       & 148          & 0.77 & 0.68      & 265      \\ 
\hline
\end{tabular}
\caption{Summary of the zoom-in simulations analysed in this work. {\bf S1} is our single star simulation, and {\bf B1, 2} and {\bf 3} are our binary star simulations. $\Delta x_{min}$ is the smallest cell size on the highest level of refinement. For system  \textbf{B1} we present results from a standard and high resolution run marked with $^\ast$ that was carried out to test for the numerical robustness of our results. The age is the final age of the oldest star (usually the primary). Masses are measured at the end of the simulation. Periods correspond to the number of binary orbits a given system made measured from the first periastron until the end of the simulation.}
\label{tab:zoom_in_sum}
\end{table}

Young stars are often found in binary or higher-order multiple systems. The number of stars found in multiple systems decreases as they approach the main sequence \citep{Chen_2013}, and multiplicity increases with primary mass \citep{Moe_2017, offner_origin_2022}. The main proposed pathways for the formation of multiple protostellar systems are turbulent core fragmentation \citep{Offner_2010,refId0} and disk fragmentation \citep{Tokovinin_2019, toomre_gravitational_1964}. In order to understand star formation and evolution, it is necessary to obtain an understanding of the formation of YSO-multiples and their interaction with their environment. 

Protostellar multiplicity has a major effect on the evolution of young stars and the proceeding formation of planets from the protoplanetary disks surrounding them \citep{jorgensen_binarity_2022}. Surveys of protostellar submillimeter continuum emission reveal protoplanetary disks of a variety of shapes, sizes, and substructures \citep{2023ApJ...951....8O,dsharp2,dsharp3}. Disk substructures in the form of rings, spirals, or cavities are hypothesised to be due to the interaction with planets formed in the protoplanetary disks \citep{2016AJ....152..222A, 2012ARA&A..50..211K, 10.1093/mnrasl/slw032}, but some substructures can arise from instabilities in the disk \citep{lesur2010subcritical,nelson2013linear}, or potentially interactions with another star \citep{cuello_flybys_2019}. 

Taking into account multiplicity effects complicates the study of star formation since it departs from the classical model of an isolated gravitationally bound and globally-collapsing core. Modelling such processes require simulations of both large scales to resolve the filamentary structures and cores, along with small scales to correctly track infall and accretion onto the protostar. Simulations that take into account large-scale ISM interaction have been able to reproduce the complex morphology of observed systems \citep{Kuffmeier_2019, jorgensen_binarity_2022}. Such simulations can realistically recreate a turbulent GMC environment where prestellar cores constantly emerge from collapsing over-densities inside the filaments through turbulent fragmentation \citep{padoan2002stellar}. The study of the formation and early evolution of individual protostars with their disks is possible through the use of sufficient adaptive mesh refinement (AMR) with grid-based simulations. \\

This is the first paper of a series focusing on Simulated Analogues. Simulated Analogues are protostars formed spontaneously in the computational model of Molecular Cloud evolution. The significance of the models of star formations carried out in this work is amplified by the ability to methodically compare the simulated systems to observations. The second paper in the series \citep{rami} describes a systematic method for matching Simulated Analogues to protostellar observations, with neural networks trained on the data analysed in this paper. In this work, we use ideal magnetohydrodynamic (MHD) simulations of \textit{ab initio} star formation to study the protostellar environment, and how disks grow in different environments. We use radiative transfer in post-processing to understand how the binarity of a system and the viewing angle affects the observed protostellar class. 

\section{Methods}

\subsection{Simulating the evolution of a GMC with \texttt{RAMSES}}

We use the code \texttt{RAMSES} \citep{RAMSES2002, RAMSESUPGRADE} to simulate star formation and early evolution inside a turbulent molecular cloud using ideal MHD, and perform `zoom-in' simulations \citep{nordlundzoomin,Kuffmeier_2017}. \texttt{RAMSES} is an oct-based adaptive mesh refinement multi-physics code. The Copenhagen version of \texttt{RAMSES} \citep{haugbolle_stellar_2018} augments the public version with modules relevant to star formation, sink particles to represent stars, support for making zoom-in models, hybrid OpenMP-MPI parallelism, improved load-balancing and pervasive vectorisation. It significantly speeds up the execution and improves the scalability of adaptive mesh refinement models with many levels of refinement making it feasible to carry long-term integration with a reasonable wall-clock time of months on 1000 CPU cores per $\approx70\kyr$ evolution of each of the zoom-in models described below.

The setup of the global simulation is a $3000\msun$ molecular cloud inside a $(4\pc)^3$ box with periodic boundaries and an isothermal equation of state with a temperature of 10 K corresponding to a sound speed of 0.18 $\kms$. Initially, we disable self-gravity and apply random solenoidal forcing at the largest scales, to mimic the impact of large-scale feedback, starting from an initial uniform density and magnetic field. The mean magnetic field strength is 7.2 $\mu$G. After 20 turnover times a fully turbulent state is created with no memory of the initial conditions and a resulting volume averaged three-dimensional velocity dispersion of $\approx1.8\kms$. At this point, self-gravity is introduced, while forcing is maintained, and the system is evolved for $2\Myr$. The model is very similar to that described in \citet{haugbolle_stellar_2018} but with globally double the linear resolution. This corresponds to a $512^3$ root grid, 6 levels of refinement, and the smallest cell size of $25\au$. It has already been used as a multi-scale model for studying core chemistry \citep{2021A&A...649A..66J, Sigurdaccepeted}, explore late infall \citep{2023EPJP..138..272K}, and as a basis for zoom-in models in a molecular cloud context \citep{jorgensen_binarity_2022}, which we explore for longer times and in more detail in this paper. Within the global simulation, sink particles form where the gas has started to gravitationally collapse. The sink particle recipe is described in \citet{haugbolle_stellar_2018}. 321 protostars are formed in the global model with a realistic initial mass function \citep{haugbolle_stellar_2018} and protostellar multiplicity distribution \citep{KURUWITA2023}. We use a Truelove criterion for refinement \citep{Truelove+97} resolving the Jeans length with a minimum of 28.8 cells at any level of refinement, except for the highest level, where sink particles form in cells reaching a density equivalent to 2 cells per Jeans length \citep{haugbolle_stellar_2018}. The number of cells is close to 300 million and the mass resolution at the highest levels is typically from $10^{-5}$ to $10^{-6}\msun$. This extraordinarily high resolution allows for a very well-resolved turbulent cascade at all scales in contrast to Lagrangian models with a fixed mass resolution and an order of magnitude less resolution elements. From the global run, we choose four isolated systems that evolve into three wide binaries and a single stellar system and perform a `zoom-in' simulation run for each of them.

Our zoom-in simulations follow the formation with a resolution down to a cell size of either $0.8\au$ or $3.1\au$ in a $10^4 \au$ radius from the primary star. We use a Truelove criterion \citep{Truelove+97} and refine on density resolving the flow with an increasing amount of cells per Jeans length reaching 64 for cells at or below $3.1\au$. The number of cells per Jeans length, $L_J$, is
\begin{equation}\label{eq:LJ}
    L_J = \frac{1}{\Delta x}\sqrt{\frac{\pi c_s^2}{G\rho}}\,,
\end{equation}
where $\Delta x$ is the cell size, $c_s$ the sound speed, $G$ the gravitational constant, and $\rho$ the density.
Outside a sphere of a radius of $1\pc$ from the centre, the resolution is at the root grid level, which in the zoom-in runs has been decreased to level 7 or cell size of $6446\au$. The ladder of refinement is given in \Cref{tab:refinement}. The high minimum $L_J$ at high resolution guarantees an adequate resolution of the disk, with typically 1 to 2 million cells per level. The large volume inside which the highest possible resolution is allowed makes it possible to resolve the pre-stellar core and follow the evolution of the binary stars. To minimise the numerical diffusion due to supersonic advection speeds, Galilean transformations into the rest frame of the primary star are continuously applied \citep{Kuffmeier_2017}. For the zoom-in simulations, we change the equation of state to use a polytropic model with varying gamma. This equation of state is derived from 1D radiation hydrodynamic simulations \citep{masunaga_radiation_2000}, and approximates some of the radiation effects during the various stages of protostellar collapse. We store outputs from the simulation for up to $260\kyr$ with a cadence of $100\yr$ allowing us to accurately track the evolution of the protostellar systems from a realistic starting condition retaining proper anchoring of the magnetic fields and the torques and inflows from the larger environment. The wide binary models are also described in \citet{jorgensen_binarity_2022}. The zoom-in runs are summarised in \Cref{tab:zoom_in_sum}.

\subsection{Dust radiative transfer}
To post-process the \texttt{RAMSES} outputs we use the dust radiative transfer code \texttt{RADMC-3D} and calculate the equilibrium dust temperature in each cell and the spectral energy distribution (SED) seen by different observers \citep{RADMC3Dmanual}. 
To limit the memory requirements and computational cost, we do not post-process the entire box but use a $30\,000\au$ cut-out centred on the primary star.
\texttt{RADMC-3D} uses a Monte Carlo method to track the propagation of photons through a dusty medium \citep{bjorkman2001radiative}. We use the same setup as in \citep{frimann2016large,2021A&A...649A..66J}, assuming a fixed gas-to-dust ratio of 1:100 and dust opacities from \citet{ossenkopf1994dust}. 
The stellar luminosity from the stars in the zoom-in run is interpolated from the stellar evolution tracks by \citet{1997MmSAI..68..807D} with a 100 kyr offset and we add the accretion luminosity to the stellar luminosity as described in \citep{2021A&A...649A..66J}. In addition, an external interstellar radiation field is included. It is similar to a Draine field \citep{1978ApJS...36..595D} with some modifications \citep{Sigurdaccepeted}. We do not apply any extinction ($A_{V,ext}=0$) to the external radiation field. This may result in slightly high dust temperatures in the outer part of the domain, but we have not found it to affect the inner part wherein the majority of the radiation is produced. When computing the SED we exclude the external radiation field and only account for photons produced in the domain.
The SED is computed using the ray-tracing capability of RADMC-3D.

To find the YSO class, the evolutionary tracer $T_{bol}$ is computed from the produced spectra as the temperature of a black body with the same mean frequency as the SED:
\begin{align}
    T_{bol} = 1.25 \cdot 10^{-11} \: 
    \frac{ \int_0^{\infty} \nu S_{\nu} \: d\nu}
    { \int_0^{\infty} S_{\nu} \: d\nu   } \: \textrm{K Hz}^{-1}\,,
    \label{eq:Tbol}
\end{align}
where $S_\nu$ is the specific intensity of the post-processed output.

\section{Results}

The zoom-in simulations are evolved until the primary stars have reached an age of up to $\approx 260\mathrm{kyr}$ to track the protostellar phase in a realistic setting.
The results are structured as follows: first, we investigate the early kinematic structure of star-forming cores in terms of the centrifugal radius profile; second, we demonstrate how the binary orbit influences the evolution of circumstellar disks and lastly, we present how projection effects can influence the observational classification.

\begin{figure}
    \centering
    \includegraphics{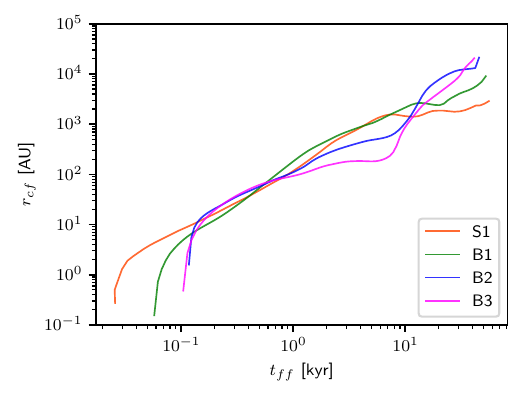}
    \caption{Centrifugal radius versus free-fall time for single and binary star simulations. The centrifugal radius is calculated using \Cref{eqn:rcf} and the time scale assume that shells collapse according to \Cref{eqn:freefall_time}.}

    \label{fig:all_ff}
\end{figure}

\begin{figure*}
    \centering
    \begin{subfigure}{0.45\textwidth}
    \includegraphics[width=\textwidth]{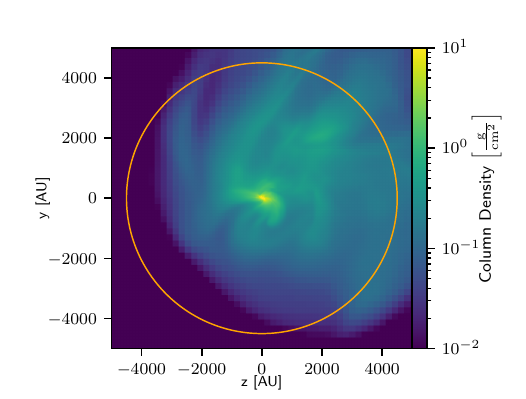}
   
    \end{subfigure}
    \begin{subfigure}{0.45\textwidth}
    \includegraphics[width=\textwidth]{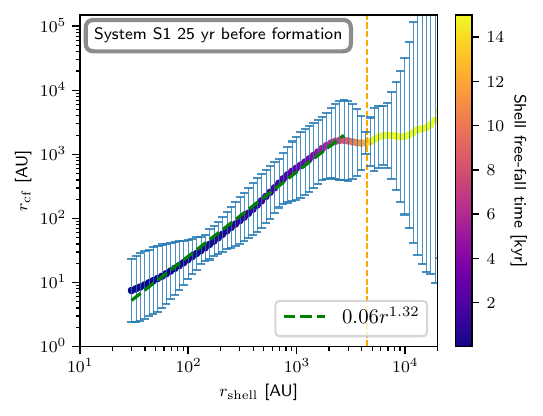}
    
    \end{subfigure}
    
    \smallskip
    
    \begin{subfigure}{0.45\textwidth}
    \includegraphics[width=\textwidth]{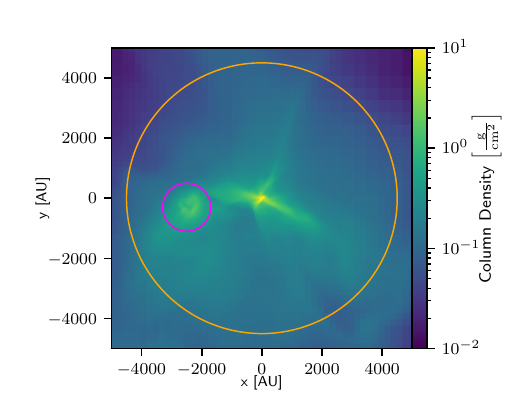}

    \end{subfigure}
    \begin{subfigure}{0.45\textwidth}
    \includegraphics[width=\textwidth]{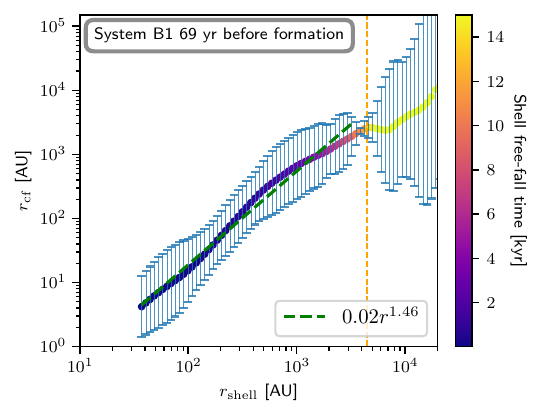}
   
    \end{subfigure}
    \caption{Integrated column density projections (left) and radial profile of centrifugal radius $r_{cf}$ (right) just before the formation of the single (upper panels) and binary (lower panels) protostellar system. Projections integrate along a box of the length of $10\,000\au$ centred at the point of collapse. The orange line represents the turnover point identified at $\sim4500\au$, and the magenta circle on the lower right panel corresponds to the clump where the secondary will form. The green dashed line in the right panels is a power-law fit to the centrifugal radius profile. The error bars indicate the standard deviation in the measured centrifugal radius in each shell.}
    
    \label{fig:enviroment}
\end{figure*}

\begin{figure*}
    \centering
    \includegraphics{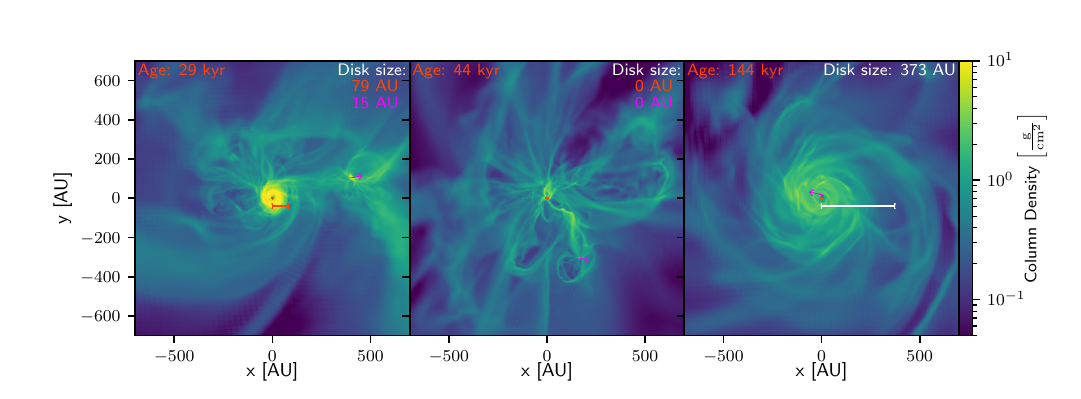}
    \caption{Gas column density of the B1 simulation calculated from a $(1400\au)^3$ cube centred on the primary star at $29\kyr$ (left panel), $44\kyr$ (middle panel), and $144\kyr$ (right panel) since the formation of the primary. It illustrates the early non-steady evolution of the proto-binary that eventually results in the formation of a close binary with a large circumbinary disk. The viewing angle is defined by the spin vector of the circumstellar disk around the primary or the circumbinary disk. The disk sizes (see \Cref{subsection:gas_dynamics}) are annotated in the upper right of each panel with red, purple and white indicating primary, secondary and circumbinary disk sizes respectively. Magenta arrows indicate the velocity of the secondary with respect to the primary and the size is proportional to the speed. Red and white bars show circumstellar and circumbinary disk sizes.}
    \label{fig:90_face}
\end{figure*}

\begin{figure}
    \centering
    \includegraphics[width= \linewidth]{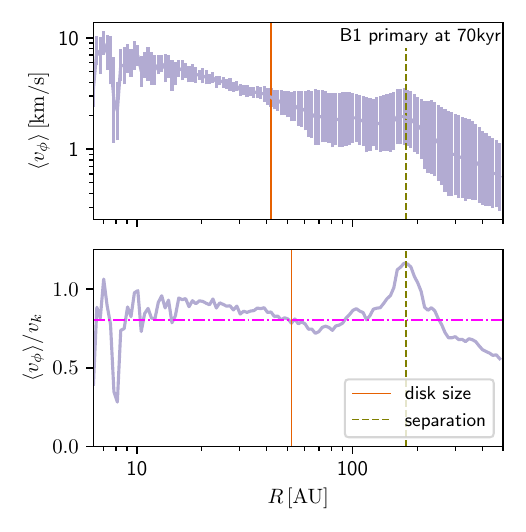}
    \caption{Circumstellar disk size measured around the primary star of system B1. \emph{Top}: angular velocity averaged in shells. The error bars indicate the spread inside each shell. \emph{Bottom}: angular velocity normalised to the Keplerian velocity. The orange line shows the estimated disk size. The vertical green line is the distance to the secondary companion. The magenta line marks the $80\%$ threshold.}
    \label{fig:90_single_disk}
\end{figure}

\begin{figure*}
    \centering
    \includegraphics[width=\textwidth]{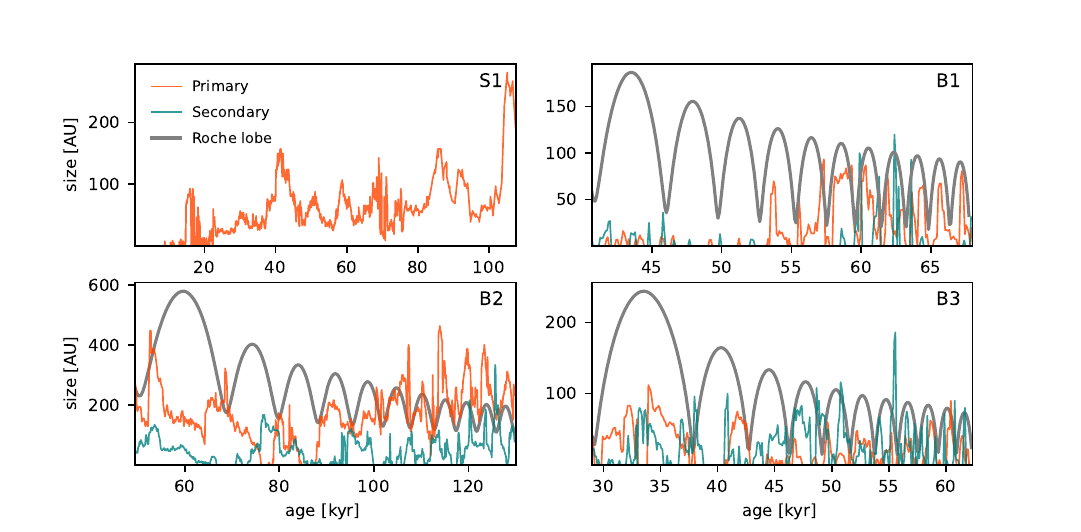}
    \caption{Roche lobe and disk size versus time. The upper left panel shows a single-star system. The other panels show measured disk sizes around the primary (orange) and secondary (blue) components for 11 orbits starting at the second periastron. The Roche lobe is shown by the thick grey line.}
    \label{fig:ds}
\end{figure*}

\subsection{Angular momentum transport, disk sizes and the centrifugal radius}
\label{ssec:centrifugal_radius}

Star-forming regions are characterised by supersonic turbulence, and the stars form in the turbulent flow where self-gravity is strong enough to overcome the thermal and non-thermal pressure, corresponding to the core environment\citep{padoan2002stellar,filementary_cores}. The random motion will always contain a residual net rotation, which is amplified by the collapse if angular momentum is conserved, leading to the formation of a flattened disk-like structure \citep{2013MNRAS.432.3320S}. Initially, the disk is highly unstable, due to the missing central mass and a propensity for gravitational instability. As the star grows, it increasingly dominates the local gravitational potential, stabilises the disk, and the rotational flow becomes Keplerian. The growth of the star is supported by infall from the surrounding core, replenishing the disk material in the process.

To investigate this picture of mass assembly and how well angular momentum is conserved in different environments, we use the last output before the star is formed to calculate the angular momentum distribution in the pre-stellar core environment for all the zoom-in simulations. As a lowest order approximation, we divide the collapsing region into concentric spherical shells and assume that each shell experiences the gravitational pull from the material interior to the shell. If the shells do not interchange angular momentum and motion is circularised, the collapse of a shell will halt when it reaches the centrifugal radius ($r_\mathrm{cf}$). At this radius, gravitational acceleration is balanced by the centrifugal force, causing matter that came from a particular shell to orbit the core in a circularised (Keplerian) orbit. The centrifugal radius, which is dependent on the specific angular momentum $j$ of the gas at a given shell, is
\begin{equation}
    r_\mathrm{cf} = \frac{j^2}{G M_<}\,,
    \label{eqn:rcf}
\end{equation}
where $M_<$, is the mass of the gas that is gravitationally acting on the shell (i.e. mass interior to the shell). Ignoring pressure forces and initial velocities, the shell will reach the centrifugal radius in a free-fall time
\begin{equation}
    {t_{\mathrm{ff}}}_{shell} = \sqrt{ \frac{9 r^3_{shell}}{128GM_<} }.
    \label{eqn:freefall_time}
\end{equation} 

We calculated the mass-weighted centrifugal radius and free-fall time in each simulation at the time step before the primary star (or star in the single star case) formed. The free-fall time was calculated for 75 shells with logarithmic spacing from $8\Delta x_{min}$ to $20\,000\au$. The results of this is shown in \Cref{fig:all_ff}. For the single star case, we see that the centrifugal radius begins to plateau for $t_{ff}\gtrsim10\kyr$ at around a centrifugal radius of $1000\au$. This implies that the specific angular momentum at larger free-fall times (i.e., larger radius) does not increase, reflecting the low turbulent environments that produce the conditions for single star formation. The centrifugal radius profile in the binary cases, however, shows an uptick at higher free-fall times, which reflects the higher angular momentum environments that promote the formation of binary and multiple stars.

\Cref{fig:enviroment} shows column density projections and the centrifugal radius as a function of radius for the S1 single and the B1 binary star systems, including error bars indicating the standard deviation of $r_\mathrm{cf}$ in each shell. The centrifugal radius profile for S1 has a smaller spread and fewer features than the binary B1, indicating that the pre-stellar core of B1 has more substructure than S1. The change in slope in the centrifugal radius marked by an orange vertical dashed line indicates a transition distance at which the gas no longer is affected by the infalling motion towards the centre of the core, and velocities are determined by the larger cloud structure. In the column density view, it can be seen this corresponds approximately to the edge of the core, which is annotated by the orange circle in the corresponding column density projection. The fitted power law, shown by the dashed green line, indicates that the spherical profile of centrifugal radius is dependent on the radius as $ \sim r^{1.4 \pm 1}$ and is consistent in both single and binary star-forming regions. For completeness, the radial profiles with column density projections for systems B2 and B3 systems can be found in the appendix in \Cref{fig:enviroment2}. 

The centrifugal radius of the prestellar environment analysis from \Cref{fig:enviroment} assumes that the gas from the environment will fall into a stable circular orbit maintaining its angular momentum and growing the disk to sizes beyond 1000 AU in less than 15 kyr. We can compare this prediction of disk growth with the measurement of disk sizes from the simulation discussed in the following section.

\subsection{Evolution of the circumstellar disks}
\label{subsection:gas_dynamics}
As the collapse progresses, a stable disk forms around the stars, however, disk growth presents differently between the single star and binary star cases. In the binary star formation simulations, we often see that circumstellar disks are regulated by the binary orbital evolution through tidal torques. This is shown in \Cref{fig:90_face} where we present density projections of the B1 simulation at different ages. From \Cref{fig:90_face}, for the binary system, we see that as it evolves the individual components form disks that are disrupted and redistributed into a circumbinary disk. While we qualitatively see these disks in \Cref{fig:90_face} and the analysis of the profile of the envelope showed that disk growth is supported, we aim to quantitatively measure these disk sizes and understand how the binarity of the system affects disk formation and evolution.

The accretion disk is the region around a sink particle that is in bulk Keplerian motion. We determine whether a given sink particle has an accretion disk and estimate the size of the disk using the gas rotational velocity profile defined with respect to the local spin vector. To calculate the spin vector, $\mathbf{J}_\mathrm{disk}$, of the gas inside an accreting disk, first, we define a $1000\au$ sphere centred on the protostars and calculate the total specific angular momentum inside it. This serves as the first approximation. Then we use that spin vector of the $1000\au$ sphere to define a $z$-axis of a disk with a radius of $150\au$ centred on the protostar. We assume a scale height of $H = 0.25 R$ to model the opening angle of the disk, therefore the height of this disk is $0.25 \times150\au = 37.5\au$. Calculating the total specific angular momentum inside that disk gives a new approximation of the spin vector. We use it as a reference to recalculate the spin inside the disk and do it recursively until convergence (usually after five iterations). With this as a reference for a cylindrical coordinate system, we calculate the mass-weighted rotational velocity in cylindrical shells. The smallest and largest shells have radii of $R=2\Delta x_{min}$ and  $500\au$ respectively. The radii of the shells are logarithmically spaced. For each cell inside a given shell, we find the velocity in a cylindrical frame of reference ($(v_r,v_\phi,v_z)$) in order to obtain the momentum of the gas inside. For each shell we can then define the mass $m_s$, mass-weighted rotational velocity $\langle v_\phi \rangle$, and spread in the mass-weighted rotational velocity $\delta v_\phi$ in a shell $s$ as
\begin{align}
    m_s &= \textstyle\sum_{i \in s} m_i \\
    \langle v_\phi \rangle & = \frac{1}{m_s} \textstyle\sum_{i \in s} v_{\phi, i}\, m_i \\
    \delta v_\phi & = \sqrt{\frac{1}{m_s} \textstyle\sum_{i \in s} \left ( v_{\phi,i} - \langle v_\phi \rangle\right)^2 m_i }\,,
    \label{eq:disvel}
\end{align}
where $\sum_{i \in s}$ is a sum over all cells, $i$, in shell $s$.

Identifying what part of the domain corresponds to a circumstellar disk in Keplerian rotation is non-trivial and somewhat ill-defined. By visually inspecting the gas distribution, we have found that the following procedure works relatively well. The edge of the disk is defined as the last shell where $\langle v_\phi \rangle > 0.8 v_K$ and $\delta v_\phi < 0.17 \langle v_\phi \rangle$. \Cref{fig:90_single_disk} shows the resulting disk size for the primary star of system B1 at $70\kyr$. In the top panel we see that at smaller radii, the spread in velocity per radial bin is lower than at larger radii. In the lower panel, we show the velocity profile in terms of the Keplerian velocity. At a radius of $\sim 50\au$, marked by the vertical orange line, the velocity profile becomes sub-Keplerian, and the spread in velocity grows significantly larger. Therefore, for this system, at this time, we find the circumstellar disk around the primary component to have a radius of $50\au$.

\subsubsection{Effect of binary interaction on the circumstellar disk size} \label{sec:TDE}

\begin{figure*}
    \centering
    \includegraphics[width=\textwidth]{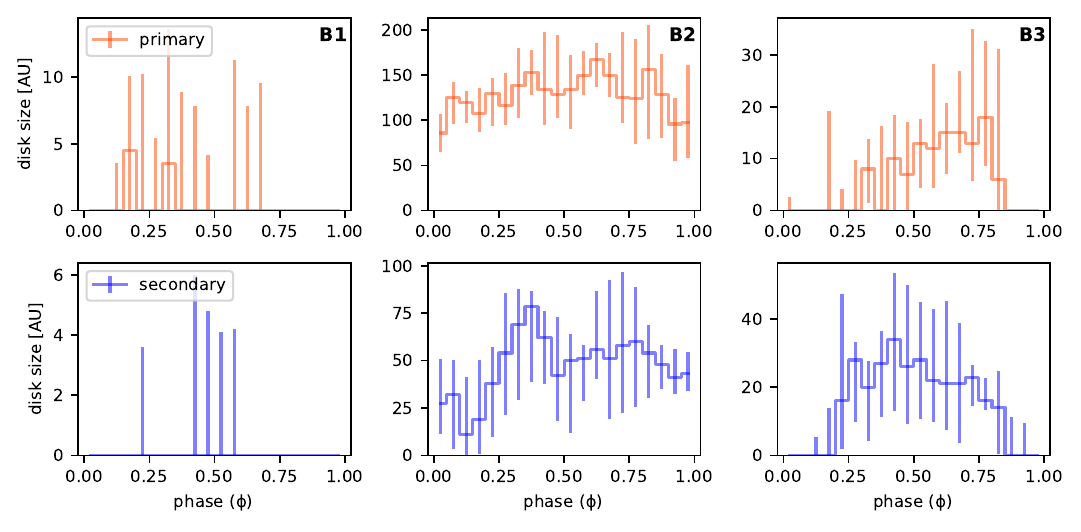}
    \caption{Phase folded median disk size from the 5th until the 15th orbit for binary systems. The upper panel (orange) shows values for the primary, while the lower panel (blue) marks the values for the secondary disk size. Each orbit is divided into 20 time bins where the median value is calculated in corresponding bins over orbits. Phase 0 and 1 correspond to periastron, while phase 0.5 is apastron.} 
    \label{fig:phase_fold}
\end{figure*}

\begin{figure*}

    \includegraphics[width=\linewidth]{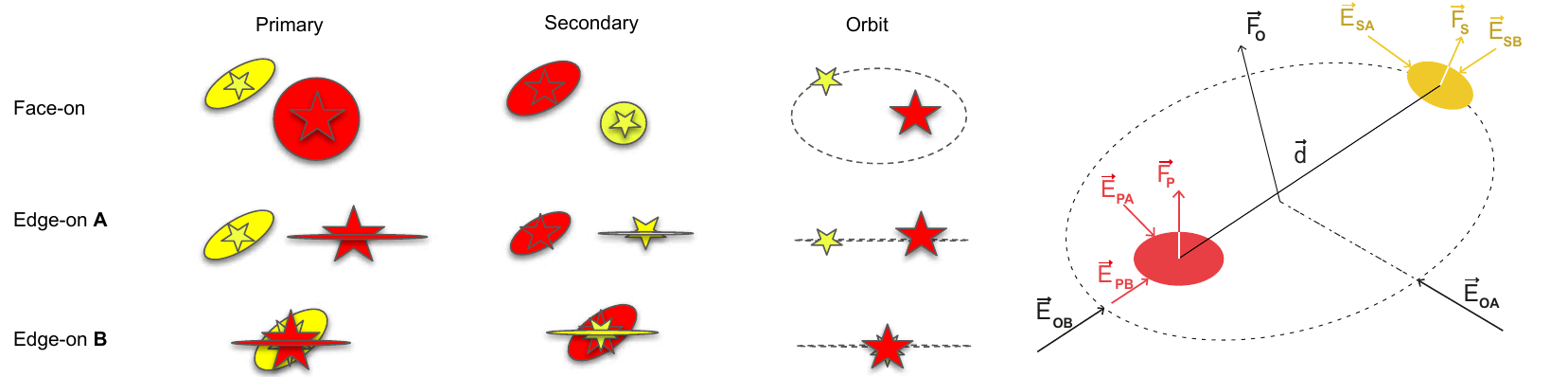}
    \caption{Schematic overview of characteristic binary protostellar system perspectives. Red and yellow colours correspond to the primary and secondary star and disk respectively, while the dashed line indicates the orbital motion. Vectors on the right panel are labelled as follows: $\protect\overrightarrow{\mathrm{d}}$ is the binary separation, $\protect\overrightarrow{\mathrm{E}}$ stands for edge-on, $\protect\overrightarrow{\mathrm{F}}$ for face-on, while subscripts P, S, O correspond to primary, secondary and orbit respectively. {\bf A} and {\bf B} edge-on perspectives are indicated by the corresponding subscript. Image Credit:  Evangelia Skoteinioti Stafyla.}
    \label{fig:tbol_schema}
\end{figure*}

To investigate how the size of circumstellar disks evolves in a binary and a single protostellar system we show a time series for the disk sizes including an outer envelope defined by the Roche lobe radii $r_L$ for the primary in the simulated systems in \Cref{fig:ds}. We calculate the Roche lobe radii of the primary as \citep{eggleton1983approximations}
\begin{equation}
r_L=\frac{0.49 q^{2 / 3}}{0.6 q^{2 / 3}+\ln \left(1+q^{1 / 3}\right)} d, \quad 0<q<\infty,
\end{equation}
where $q$ is the mass ratio and $d$ is the binary separation. The Roche lobe yields the maximum radius at which bound material can exist around each of the stars. It should be a good proxy for the maximum stable size of individual circumstellar disks.

For the single star system S1, at times before $\sim15\kyr$ the analysis estimates that the disk size is smaller than the minimum resolution ($0.8\au$). After $\sim15\kyr$ the disk size grows to somewhere between $30\au$ and $100\au$. It is modulated by environmental factors, such as large-scale accretion events, and the stability of the disk itself, but rarely drops below the resolution scale except for a major event a $\sim70\kyr$. This intermittency in the early phases of disk evolution connected to the infall of material has a period of $\sim10kyr$ which compares well to the observed frequency of episodic accretion events as traced by CO-sublimation \citep{2017A&A...602A.120F,2019ApJ...884..149H}.

In binary systems, disk sizes often suddenly drop below the minimum resolution. When looking at the disk evolution in the binary case it can be seen that the periodic growth and dispersal of disk size is modulated by the Roche lobe radii corresponding to the orbital evolution. To investigate the dependence of the disk size on the binary orbit we phase-folded the disk size evolution. The time in each orbit is divided into 20 time bins so that there are 10 bins between each periastron and apastron. In each time bin, we find the median disk size and then phase folding is done by taking the median of the values in the corresponding bin across orbits. The orbit-to-orbit spread in each bin is taken to be the distance from the median to the end of the range defined by one standard deviation below/above the mean. Resulting phase-folded disk size evolution for 10 orbits after the 5th periastron is shown in \Cref{fig:phase_fold}. The left and the right panel (systems B1 and B3) of \Cref{fig:phase_fold} show that the disk size around both the primary and secondary are generally the largest around the phase of 0.5 corresponding to apastron. The same trend is not seen in system B2. We believe this is because the disk size is overestimated by a corotating envelope, making it indistinguishable from the disk when looking at the Keplerian velocity profile. This can be seen on the lower left panel of \Cref{fig:ds} at times larger than 100 kyr, where at some points the disk size grows larger than the Roche lobe which is impossible for the circumstellar disks. The co-rotating envelope is illustrated in \Cref{fig:49_extended}.

\begin{figure*}
 
    \includegraphics[width= 0.7\textwidth]{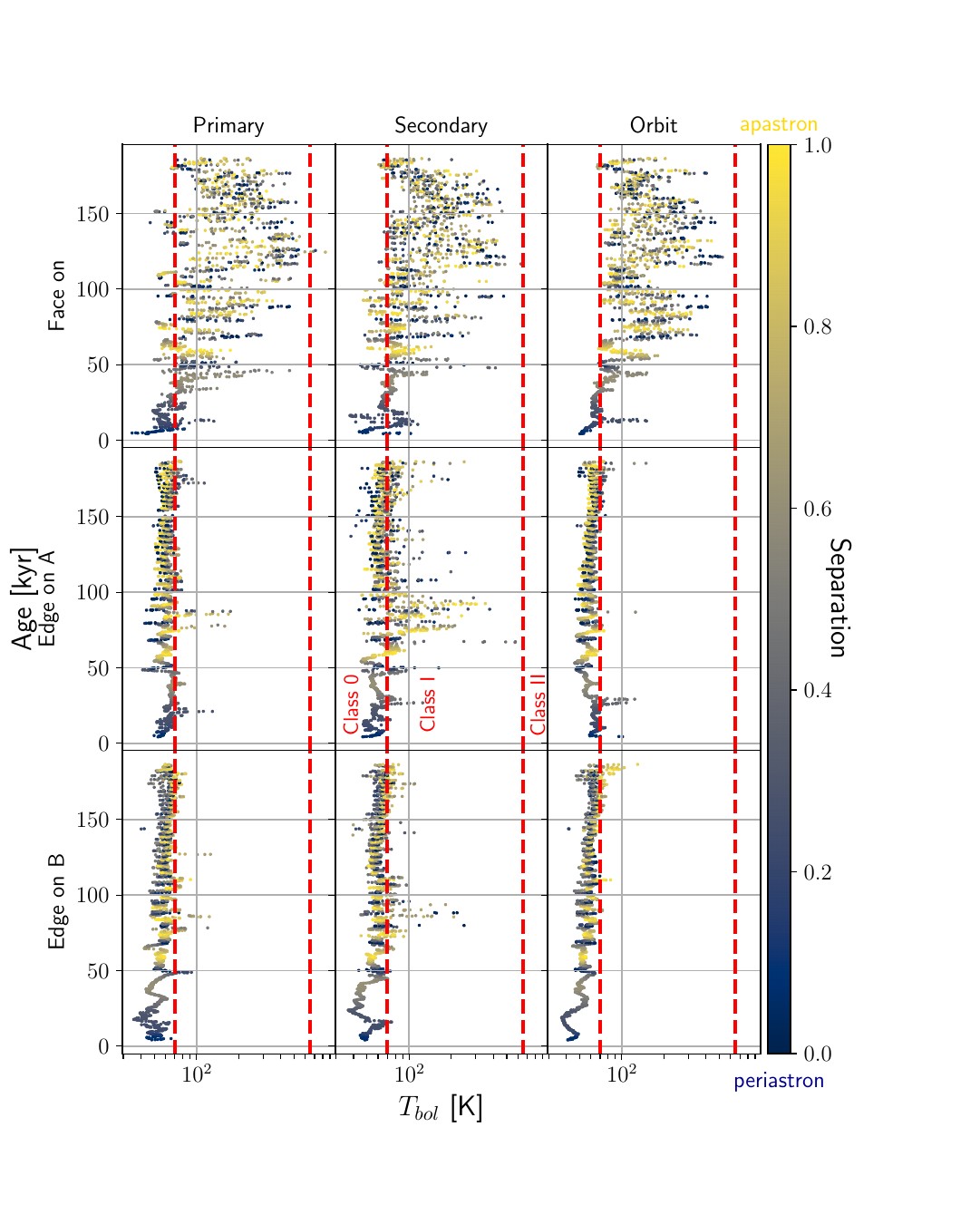}

\caption{Bolometric temperature evolution for the B2 system. The vertical dashed red lines indicate the protostellar class thresholds, as defined in \Cref{eqn:ysoclass}. The top row represents face-on viewing angles (with normal vectors $\vec{F_p}$, $\vec{F_s}$, $\vec{F_o}$ from left to right). The middle row corresponds edge-on angles normal to the separation $\vec{d}$ (with normal vectors $\vec{E_{PA}}$, $\vec{E_{SA}}$, $\vec{E_{OA}}$ from left to right). Finally, the bottom row shows edge-on angles parallel to the separation $\vec{d}$ (with normal vectors $\vec{E_{PB}}$, $\vec{E_{SB}}$, $\vec{E_{OB}}$ from left to right).}
    \label{fig:Tbol_char_49}
\end{figure*}

\subsection{Bolometric temperature evolution from different viewing angles}
\label{ssec:bolometric_temperature}

The bolometric temperature $T_{bol}$ can be used to classify of YSOs. This has traditionally been related to the degree of embeddedness and the evolutionary stage of the YSO, but given that accretion an-isotropic and outflows can carve cavities, $T_{bol}$ is affected by the line of sight. In the case of binary YSO systems, this is even more complex and the structure of their environment is highly asymmetric and time-dependent. In this section, we investigate how binarity and projection effects affect the observed bolometric temperature of the system and hence the classification.

Several lines of sight can be defined based on the geometry of the binary protostellar systems. Because of the higher column density, we expect disks to be effective in absorbing and re-emitting radiation at lower temperatures, while cavities carve a low optical depth pathway for the photons. To explore the full variability that can be expected for different observers we define nine different points of view. We use the disk spin vector $\mathbf{J}_\mathrm{disk}$ obtained in \Cref{subsection:gas_dynamics} to define a line of sight which produces a \textbf{face-on} projection of the system. The \textbf{edge-on} view, which is perpendicular to $\mathbf{J}_\mathrm{disk}$, has a $2\pi$ degeneracy. But given a binary system, this degeneracy is broken. We choose the two characteristic edge-on views: \textbf{Edge on A} is \textit{orthogonal} to the separation vector between the two stars, and \textbf{Edge on B} is where both binary companions get projected to the \textit{same point} along the separation vector. This gives three orthogonal observers for each star. Furthermore, a similar set of observers can be chosen but based on the orbital spin vector. These nine binary system perspectives defined by these lines of sight are illustrated in \Cref{fig:tbol_schema}.

\begin{figure*}

    \includegraphics[width= 0.7\textwidth]{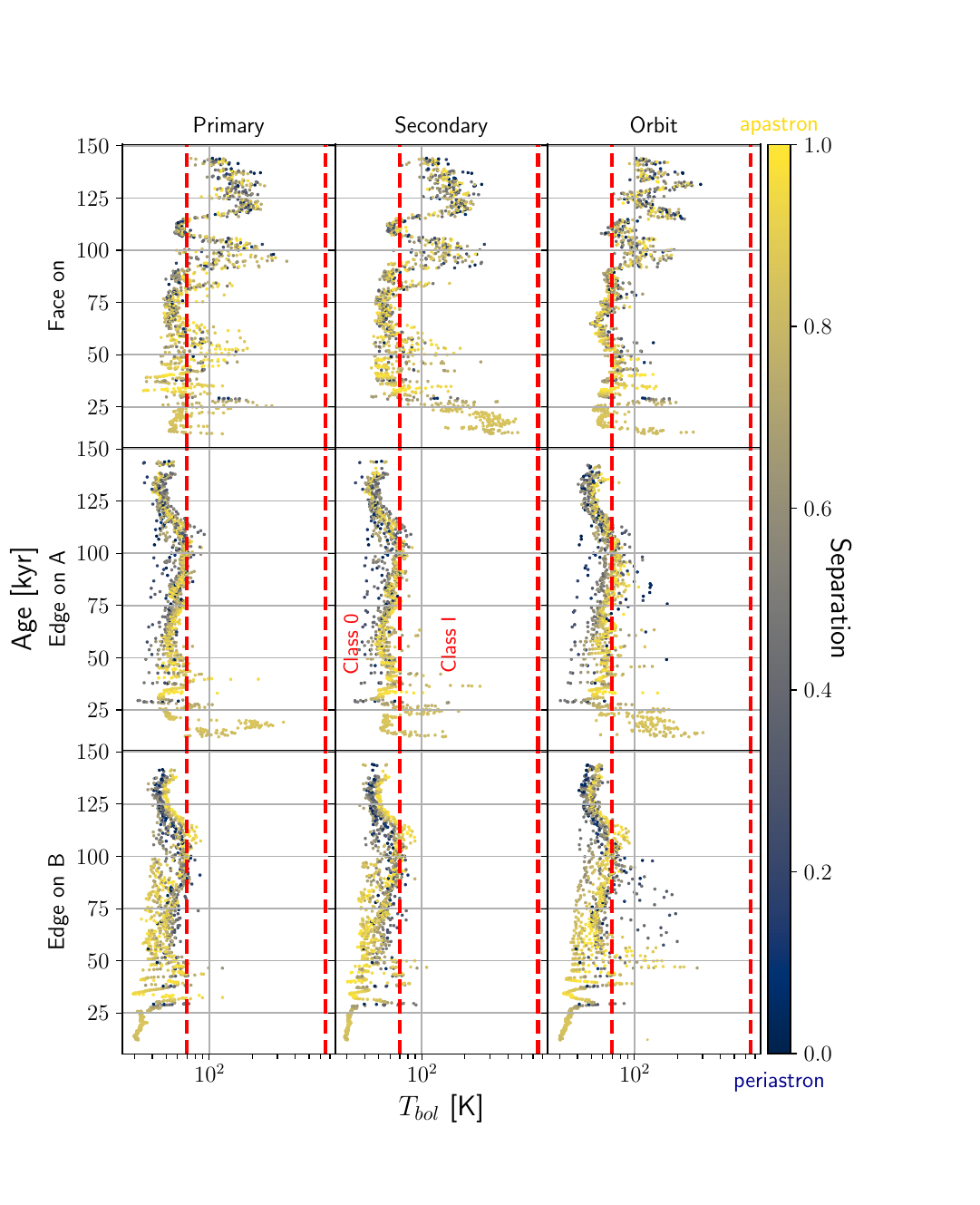}
    \caption{Same as \Cref{fig:Tbol_char_49} but for system B3 }
    \label{fig:Tbol_char_164}
\end{figure*}

\begin{figure*}

    \includegraphics[width= 0.7\textwidth]{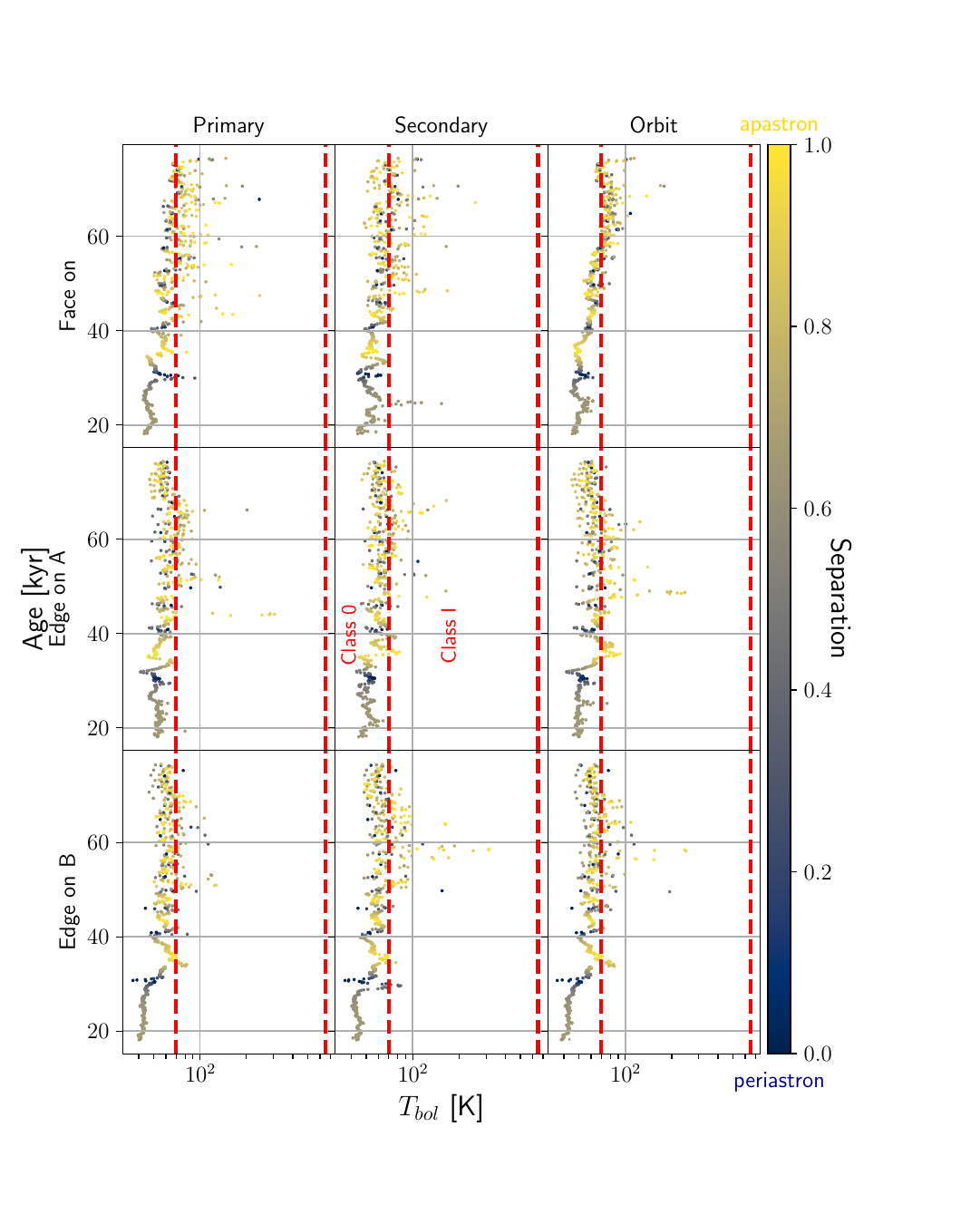}
    \caption{Same as \Cref{fig:Tbol_char_49} but for the high resolution run B1$^\ast$  }
    \label{fig:Tbol_char_90}
\end{figure*}

At each \texttt{RAMSES} time step synthetic SEDs and bolometric temperatures were calculated for each of these viewing angles using \texttt{RADMC-3D} and the resulting time-series is shown in \Cref{fig:Tbol_char_49} for B2, \Cref{fig:Tbol_char_164} for B3, and \Cref{fig:Tbol_char_90} for B1. 

These figures show that face-on projections give rise to a more evolved class (higher $T_{bol}$ - less obscuration) because of the lower optical depth due to outflows, and they have significantly higher variability. It takes time to develop the cavity and the first tens of kyr of the systems are classified as class 0 from all perspectives.

There is a general agreement in classification when looking at the system from different edge-on perspectives but face-on gives different classifications. Face-on view gives a classification of Class I after 50 kyr in the evolution at \Cref{fig:Tbol_char_49} while edge-on perspectives give a consistent classification of Class 0. This is because the edge-on perspective is optically thick giving a classification of a more embedded object than it actually is in later stages. 

In \Cref{fig:Tbol_char_49} we also see significant variation in $T_{bol}$ over the course of an orbit. This is the strongest in the face-on views (top row), where we see as the system approaches the periastron, indicated by the data points becoming bluer, the $T_{bol}$ increases, making the system look more evolved. The $T_{bol}$ decreases as the system approaches apastron, indicated by the data points becoming yellow, making the system look less evolved. At periastron, the system experience disk disruption events induced by episodic accretion and tidal interaction discussed in \Cref{sec:TDE}. In this case, the apparent protostellar class is sensitive to the binary orbit.

Long-term variability is observed in \Cref{fig:Tbol_char_164}. The $T_{bol}$ varies on the order $\sim 1000$K seen in the face-on row of \Cref{fig:Tbol_char_164} after $100$ kyr of evolution. This is attributed to inflows and outflows of material near the system since the time scale of these fluctuations is larger than the time of the orbit. Despite the long term variability, we also see variation over an orbit, but not as dramatic as in the B2 case. Other panels corresponding to the edge-on view do not show a correlation with the binary orbit indicated by the constant variability in $T_{bol}$ value across orbits. 

The B1 system shown in \Cref{fig:Tbol_char_90} is relatively stable compared to the other binary systems. However, the $T_{bol}$ of this system was only calculated up to $80\kyr$. This system has the lowest mass of all the binaries. 

\section{Discussion}
\label{sec:discussion}
\vspace{-0.2cm}
\subsection{Gas dynamics in protostellar binaries}
\label{ssec:discussion_disks}
\vspace{-0.1cm}
Centrifugal radius is often used as an estimation of the evolution of the disk size \citep{krasnopolsky2002self,dullemond2006accretion}. Results for the binary (B1) and single star-forming regions, shown in \Cref{fig:enviroment} and \Cref{fig:enviroment2} (right panel) both agree that if the envelope is purely gravitationally free-falling onto the central object, disk size would be monotonically increasing up to the sizes of $\sim10\,000\au$. Comparing those predictions to the actual disk size in \Cref{fig:ds}, we see that the disk size is not rising monotonically and that the measured disk size does not go above $200\au$ in any of our simulated systems throughout their evolution of $150\kyr$ which is beyond any of the calculated free-fall times. These findings imply that the centrifugal radius of material at early times is a poor tracer of disk size, and that angular momentum transport happens at both core and disk scales. This is supported by the results of \citet{kuffmeier2023rejuvenating} who used a complementary approach based on the accreted passive tracers and find the same order of magnitude ($\sim10\,000\au$) as in our model. It is in contrast to the conclusions of \citet{2020A&A...637A..92G}, which estimated the projected specific angular momentum for class 0 sources in the CALYPSO survey. The differences can be attributed partly to uncertainties in measuring the central mass, and partly because they consider gas much closer to the disk. As we have seen in our results, the inner protostellar envelope spins up over the first tens of kyr, and therefore it can be expected to yield a centrifugal radius close to the disk size. This does not, however, imply that angular momentum is conserved throughout the envelope. 

During the binary evolution, the circumstellar disks are regularly disrupted and have an asymmetrical structure attributed to the influence of the binary orbit, similar to what has been seen in observations and other models \citep{ragusa2017origin,muto2015significant}. The temporal fluctuations and sudden drops below the resolution scale in disk size in our models indicate that the binary orbit transfers angular momentum to the surrounding gas through a cycle of tidal disk disruption \citep{kuruwita_dependence_2020}. At apastron, the disk growth is governed by infall from the envelope and fall-back of material, but as the binary approaches periastron, tidal forces and dynamical drag accelerate core material close to the disks and launch spiral waves through circumstellar disks with the exchange of angular momentum, enhanced accretion and expansion of the disk. The end result is that matter is lost from the disk by accretion, outflows, and shedding. When the system approaches the apastron, a fraction of the shed material falls back to a stable orbit but at a larger radius, completing the cycle. As the systems progress through many orbits, periodic dispersion and subsequent growth of disk size lead to the creation of large circumbinary disks through the extraction of angular momentum of the binary orbit to the surrounding gas as illustrated in \Cref{fig:90_face}. Column density projections show how the systems at earlier times have two distinct circumstellar components while the same system at the later stage develops a single circumbinary component. In the middle panel, a variety of cavities produced by these episodic accretion bursts are clearly visible. The gas bridge connecting the two stars at early times is a co-rotating left-over of this process. This evolution is believed to be seen in the observations of a young binary by \citet{diaz2022physical}. Figure 9 from \citet{diaz2022physical} shows a similar spiral binary structure as seen in \Cref{fig:90_face} -  three spiral features were identified in 0.9 mm ALMA continuum emission image of SVS 13 protobinary system. 

An important implication of the evolution seen in our models is that it takes more than 20 orbits to build up the circumbinary disk in the case of a wide binary, and therefore we expect these systems to be more evolved with an age of at least $100\kyr$.

The observed cavities and a circumbinary structure seen in \Cref{fig:90_face} have also been seen in much more idealised models of evolved systems where the circumbinary disk is a prescribed result of the setup \citep[see e.g.][]{matsumoto2019structure,kuruwita_role_2019}. 

While the overall implications for the dynamics and evolution of protostellar binaries are clear, a small quantitative caveat is the lack of a robust procedure to estimate the disk sizes using the method described in \Cref{subsection:gas_dynamics}. The method has some difficulty distinguishing between the disk and a Keplerian corotating envelope. We attribute this to a lack of a density threshold. Values higher than the Roche lobe radius in the lower left panel of \Cref{fig:ds} show that the method gives individual estimates for the disk size even when the gas is not gravitationally bound to that specific central object. Specifically, system B2 has a highly corotating envelope which gives an overestimated disk size after 120 kyr in \Cref{fig:ds}. The corotating envelope and the binary orbit create an extended disk-like structure that is rotating near Keplerian speed which interferes with the calculation of disk size based on the Keplerian velocity profile. The difficulty in distinguishing the disk from the surrounding corotating envelope is illustrated in \Cref{fig:49_extended}.

\vspace{-0.2cm}
\subsection{Impact of binarity and viewing angle on the apparent evolution.}
\label{ssec:discussion_synth_obs}

Inferring age from protostellar observations and characterisation of their evolution is a challenging task due to obscuration coming from the turbulent nature of star-forming environments. For single stars, it is known that the different tracers of age do not always agree \cite{evans2009spitzer} and various observational uncertainties such as projection effects can also cause differences in the categorisation of the state of evolution \citep{frimann2016large}. Protostellar multiplicity adds another layer of complexity in determining the evolutionary stage.

In \Cref{ssec:bolometric_temperature} we investigated how the binarity and the viewing angle affect the apparent evolutionary stage of the simulated systems. Results shown in \Cref{fig:Tbol_char_49}, \ref{fig:Tbol_char_164}, and \ref{fig:Tbol_char_90}, show that $T_{bol}$ is not a monotonic function of time, similar to what was found by \citet{frimann2016large}.

Comparing the results, we see that there is no trend in the evolution of bolometric temperature across systems and that different binary systems have a varying $T_{bol}$ evolution, especially when viewed face-on. The strongest face-on variation over an orbit is seen in our most massive binary B2 in \Cref{fig:Tbol_char_49}. The variation seen edge-on is minimal, except for the secondary, which probably is a consequence of the secondary spin axis not aligning with the primary and orbital spin axes. For the edge-on view of B2, the system mostly remains with a Class 0 classification, but for the low-mass binaries (B3 in \Cref{fig:Tbol_char_164} and B1* in \Cref{fig:Tbol_char_90}) with the edge-on viewing angle, the systems move between Class 0 and I more frequently.

The stability in protostellar classification for the edge-on view in the B2 simulation may be because larger, denser, more stable disks are formed in this massive system. Whereas in the lower mass binaries, the disks are disrupted more strongly by the binary interactions, leading to stronger optical depth variation.

Bolometric temperatures from the line of sight corresponding to halfway between the face-on and edge-on are shown in \Cref{fig:Tbol_45_49} and its profile is more consistent with the face-on perspective for the corresponding system from \Cref{fig:Tbol_char_49}.

The upper left panel of \Cref{fig:90_face} shows a transient bridge-like structure of the gas connecting the individual binary components, as previously seen in \citet{Kuffmeier_2019}. The differences between the A and B points of view (perpendicular and parallel to the separation vector) could be due to the obscuration by the gas bridge and individual disks.

$T_{bol}$ is sensitive to viewing angle, and observed apparent age differences between secondary and primary stars may be due to column depth differences, in particular, if the gas bridge is in the line of sight. Out-bursting protostellar binaries could masquerade as later-type stars, approaching class II. Therefore, it is important to include complementary evolutionary tracers, such as column density, and chemistry \citep{2020ARA&A..58..727J}. Our results indicate that a number of known Class I and Class II objects may be much younger than what is inferred from their bolometric temperature. It could explain the difference in protostellar classes observed in some binary systems such as e.g.,~VLA 1623 \citep{2023MNRAS.522.2384M} to be a combined result of viewing angle and amount of material in the inner core.

\vspace{-0.3cm}
\section{Conclusions}
\vspace{-0.1cm}
In this paper, we investigated how binarity affects the physical and apparent protostellar evolution, using ab initio zoom-in ideal MHD simulations embedded in a larger molecular cloud context and radiative transfer post-processing. To our knowledge, these models are the first to follow binary formation with millions of cells at AU-scale resolution and in a full molecular cloud context without ad hoc or isolated initial conditions for enough time to allow circumbinary disks to arise serendipitously.

Our main results are:

\begin{enumerate}
    \item The angular momentum budget at the core scale could support massive disks ($\sim 10\,000\au$), but due to angular momentum transport at all scales, actual disk sizes are orders of magnitude smaller. This severely limits the utility of using the centrifugal radius as a proxy for the disk size.
    \item Binarity influences disks through periodic disruptive events that destroy circumstellar disks and eventually form circumbinary disks. Disk material gets redistributed predominantly at periastron crossing due to the tidal forces of the companion, and the disk typically grows close to the apastron.  
    \item Bolometric temperature as a measure of obscuration is sensitive to projection effects due to the spatial distribution of gas surrounding deeply embedded protostellar systems.
    \item Classification based on the $T_{bol}$ is not monotonic as simulated systems can go from Class I back to Class 0. Associating the YSO class and the evolutionary stage is non-trivial.
    \item In proto-binaries we find high variability on short times-scales in $T_{bol}$ mostly in phase with the orbital periods. Systems can even cross a class boundary multiple times during one orbit.
    \item Classification is influenced by the inflows from the larger scale environment, which can make the system appear more embedded if they cross the line of sight. 
\end{enumerate}

The methodology applied in this paper naturally generates a rich metadata description for simulation outputs in star-forming simulations. These data can be used to match the models to specific observations giving complementary insight into the kinematics and evolution of the system \citep{REGGIE}. The process is very labour-intensive and requires the hand-selection of matching simulation outputs. In a companion paper \citet{rami} we explore how to draw robust conclusions from selection outputs, address the human bias, and automatise the process of matching simulations and observations using machine learning.

\vspace{-0.3cm}
\section*{Acknowledgements}
\vspace{-0.1cm}
The research leading to these results has received funding from the Independent Research Fund Denmark through grant No.~DFF 8021-00350B (TH,RLK). The project has received funding from the European Union’s Horizon 2020 research and innovation program under the Marie Sklodowska-Curie grant agreement No. 847523 ‘INTERACTIONS’ (RLK). The astrophysics HPC facility at the University of Copenhagen, supported by research grants from the Carlsberg, Novo, and Villum foundations, was used for carrying out the simulations, the analysis, and the long-term storage of the results. RLK also acknowledge funding from the Klaus Tschira Foundation. \texttt{yt} \citep{turk_yt:2011}, \texttt{MATPLOTLIB} \citep{Hunter:2007} and \texttt{NumPy} \citep{harris2020array} were used to visualise and analyse these simulations. Special thanks to Evangelia Skoteinioti Stafyla for illustrating \Cref{fig:tbol_schema}.

\vspace{-0.3cm}
\section*{Data Availability}
\vspace{-0.1cm}
All data and tools are available from the authors upon reasonable request. 

\vspace{-0.3cm}
\bibliographystyle{mnras}
\bibliography{references}
\appendix

\vspace{-0.4cm}
\section{Stellar nursery gas dynamics}

\begin{figure*}
    \centering
    \begin{subfigure}{0.45\textwidth}
    \includegraphics[width=\textwidth]{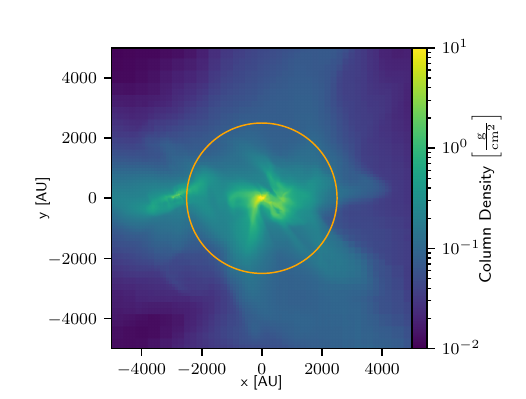}
 
    \end{subfigure}
    \begin{subfigure}{0.45\textwidth}
    \includegraphics[width=\textwidth]{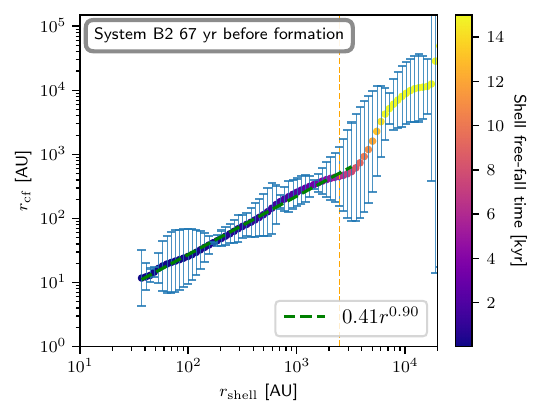}
   
    \end{subfigure}
    
    \smallskip

    \begin{subfigure}{0.45\textwidth}
    \includegraphics[width=\textwidth]{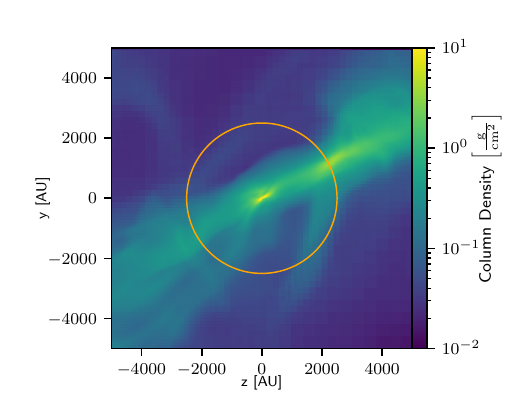}
 
    \end{subfigure}
    \begin{subfigure}{0.45\textwidth}
    \includegraphics[width=\textwidth]{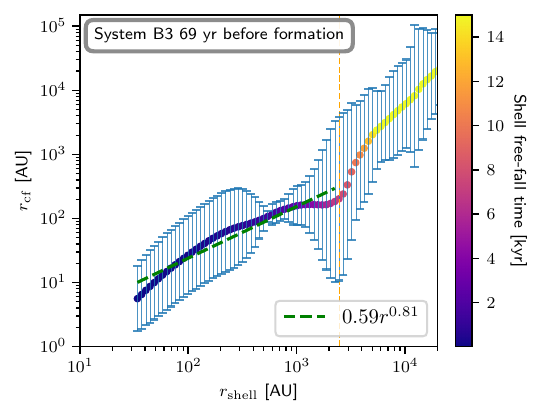}
   
    \end{subfigure}

   \caption{Integrated column density projections (left) and radial profile of centrifugal radius $r_{cf}$ (right) just before the primary component formation of the B2 and B3 simulations. Projections are integrated along a box of the length of $10\,000\au$ centred at the point of collapse. The orange line shows the turnover point identified at 2500 AU from the point of collapse of the primary, For both systems this line intersects the clump of gas where the secondary will form.}

    \label{fig:enviroment2}
\end{figure*}

\Cref{fig:enviroment2} reproduces the bottom panel of \Cref{fig:enviroment} for the B2 and B3 simulations. These systems have shallower Centrifugal radius profiles, which may indicate that these cores have lower angular momentum. The turnover points in their profiles correspond to the location where the secondary companion will form.

\begin{figure*}
    \centering
    \includegraphics{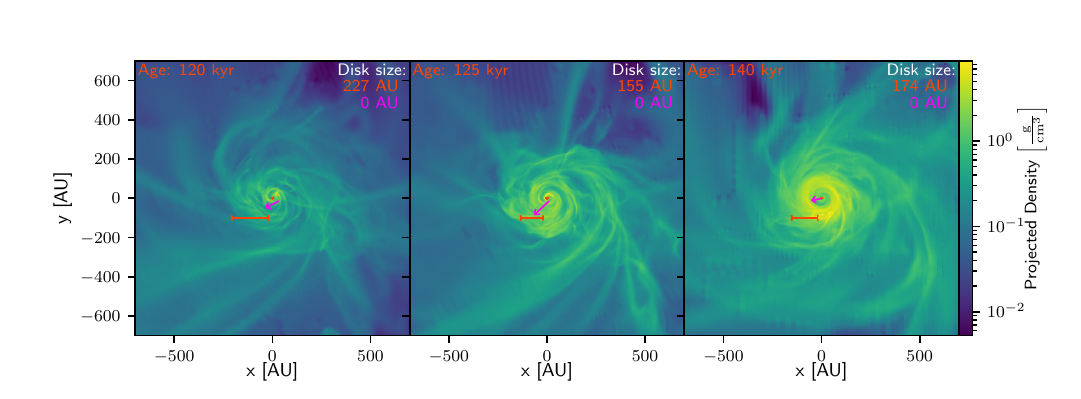}
     \caption{Gas column density of the B2 binary system calculated from a $(1400\au)^3$ cube centred on the primary star at $120\kyr$ (left panel), $125\kyr$ (middle panel), and $140\kyr$ (right panel) since the formation of the primary. The calculated disk size is indicated in red (purple) text for the primary (secondary) disk in the upper right of each panel. Magenta arrows indicate the velocity direction of the secondary in the primary rest frame and the size is proportional to the speed. Red bars show circumstellar disk sizes, measured using the method described in \Cref{subsection:gas_dynamics}. The view is face-on with respect to the primary star. }
    
    \label{fig:49_extended}
\end{figure*}

\Cref{fig:49_extended} illustrates the co-rotating envelope around a simulated binary system B2. The envelope rotating in a Keplerian fashion hinders the disk detection method from distinguishing the edge of the accretion disk based on the Keplerian velocity profile. A way to improve the method of determining disk size would be to incorporate the surface density in the calculation to differentiate between the disk and the envelope.

\onecolumn
\begin{multicols}{2}
\section{Apparent evolution from non-characteristic viewing angles} 

\begin{figure*}
    
\includegraphics[width=0.7\textwidth]{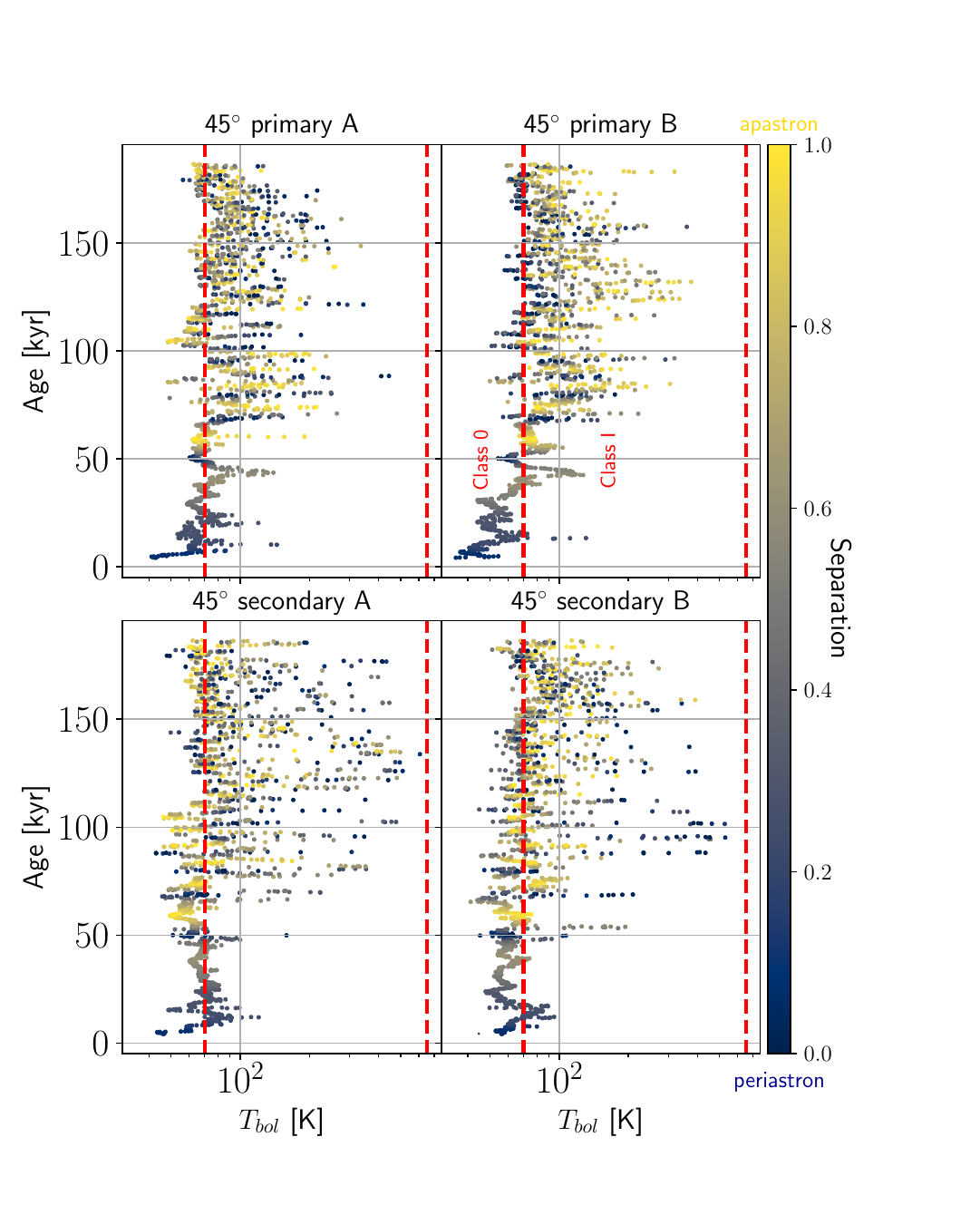}
    \caption{Bolometric temperature evolution for the B2 system seen at an off-axis angle of $45^\circ$. The viewing angles are defined as the sum of the face-on unit vectors $\vec{F}_{p,s}$ and the edge-on unit vectors $\vec{E}_{PA,SA}$ (left column) and $\vec{E}_{PB,SB}$ (right column).}
    \label{fig:Tbol_45_49}
    \vspace{-0.5cm}
\end{figure*}

\Cref{fig:Tbol_45_49} shows the bolometric temperature evolution for the B2 simulation from a point of view corresponding to the angle that is halfway in-between edge on and face-on primary. Point of view A represents the point of view where the binary components are seen next to each other, while in the B point of view, the secondary gets projected on the same point as the primary, as illustrated in \Cref{fig:tbol_schema}. The bolometric temperature at these points of view is most similar to the face-on view of the corresponding systems from \Cref{fig:Tbol_char_49}. This is expected, because most of the optical depth difference between face-on and edge-on is generated in the disk plane, and the disk opening angle is less than $45^\circ$. However, the envelope material is also concentrated towards the orbital midplane which in this case coincides approximately with the disk planes, and while there is an overall resemblance to the face-on view, in general, the synthetic bolometric temperatures are comparatively lower.
\end{multicols}

\bsp	% typesetting comment
\label{lastpage}
\end{document}